\documentclass[twocolumn,preprintnumbers,superscriptaddress,endnote,nofootinbib,prd]{revtex4}

\pdfoutput=1

\usepackage{graphicx}
\usepackage{latexsym}
\usepackage{amsfonts}
\usepackage{amssymb}
\usepackage{amsmath}
\usepackage{slashed}
\usepackage{array}
\usepackage{dcolumn}
\usepackage{feynmp}
\usepackage{comment}
\usepackage{xcolor}

\definecolor{red}{rgb}{1.0, 0, 0}
\definecolor{green}{rgb}{0.0, 1.0, 0.0}

\newcommand{\tr}{\text{Tr}}
\newcommand{\eps}{\varepsilon}
\newcommand{\D}{\mathcal{D}}
\newcommand{\Lag}{\mathcal{L}}

\begin{document}
\title{Non-perturbative proton stability}

\author{Adam Martin}
\affiliation{Theoretical Physics Department, Fermilab, Batavia, IL 60510}
\author{Gerben C. Stavenga}
\affiliation{Theoretical Physics Department, Fermilab, Batavia, IL 60510}
\preprint{FERMILAB-PUB-11-529-T}
\date{\today}

\begin{abstract}
Proton decay is a generic prediction of GUT models and is therefore an important channel to detect the existence of unification or to set limits on GUT models. Current bounds on the proton lifetime are around $10^{33}$ years, which sets stringent limits on the GUT scale. These limits are obtained under `reasonable' assumptions about the size of the hadronic matrix elements. In this paper we present a non-perturbative calculation of the hadronic matrix elements within the chiral bag model of the proton. We argue that there is an exponential suppression of the matrix elements, due to non-perturbative QCD, that stifles proton decay by orders of magnitude -- potentially $\mathcal O(10^{-10})$. This suppression is present for small quark masses and is due to the chiral symmetry breaking of QCD. Such a suppression has clear implications for GUT models and could resuscitate several scenarios.
\end{abstract}

\maketitle

\section{Introduction}
\label{sec:intro}

Proton decay is an important prediction of numerous models of beyond the standard model (SM) physics.  The decay is induced by higher-dimensional, baryon-number violating operators, suppressed by some high scale, $M_{GUT}$~\cite{Georgi:1974sy}. Although there are considerable uncertainties in the strength of the baryon violating couplings, they are usually assumed  to be $O(1)$. One can then make an order-of-magnitude estimate of the proton lifetime. For example, a $p\rightarrow \pi\,+\,\ell$ decay channel contributes
\begin{equation}
\Gamma_p=2\pi\, |\Lambda|^2\, |\langle\pi|O|p\rangle|^2 \rho(m_p),
\label{eq:pertmatelem}
\end{equation}
where $|\Lambda|^2$ contains the non-hadronic part of the matrix element along with any perturbative coefficients, e.g. $\Lambda \sim 1/M^2_{GUT}$ when baryon-number violation is caused by four-fermion interactions. For a more complete estimate of Eq.~(\ref{eq:pertmatelem}), one also needs to know the hadronic matrix elements, which have previously been estimated in lattice calculations~\cite{Tsutsui:2004qc, Aoki:2008ku, Cooney}, chiral lagrangians~\cite{Claudson:1981gh} and other approaches. In this paper, we calculate the hadronic matrix elements within the chiral bag model of the proton~\cite{Jackiw:1975fn, Vento:1980mu, Goldstone:1981kk, Goldstone:1983tu, Brown:1984sx, Mulders:1984df, Vepstas:1984sw, Vepstas:1990tw}. The chiral bag model is the marriage of two interesting phenomenological descriptions of protons. At short distances the proton is described as a `bag' of free fermions -- massless, or nearly massless fermions in a spherical bag subject to physically motivated boundary conditions. Outside of the bag radius, the proton is described with a Skyrme~\cite{Skyrme:1961vq, Skyrme:1962vh} model.

 Combining the two descriptions provides an improved model of the proton. Unlike the original bag models (the so-called MIT bag~\cite{ Chodos:1974je,  DeGrand:1975cf, Inoue:1975bk, Callan:1978bm,  Brown:1979ui, Brown:1979ij}), many physical results in the chiral bag description are insensitive to the bag radius~\cite{Nadkarni:1984eg, Rho:1994ni,DeFrancia:1995xi}. Simply, the smaller (larger) the bag, the more the proton is carried in the Skyrmion (bag). If the bag radius is taken completely to zero, we recover a purely Skyrmionic description of the proton. The Skyrme model has had considerable success in describing some baryonic properties~\cite{Skyrme:1961vq, Skyrme:1962vh,Witten:1983tx, Witten:1983tw}, however it is inadequate for describing proton decay since baryon number is identified with the winding of the pion field into the Skyrmion and is therefore topologically conserved. However, for a Skyrmion with a `hole', topological conservation is not exact and the configuration can be un-wound, a process which can be interpreted as the Skyrmion (proton) decaying to the topologically trivial ground state. The intuitive expectation is that the unwinding of a Skyrmion involves tunneling through a potential barrier and therefore comes with some exponential suppression. The smaller the hole, the harder it is for the proton to shed its topological portion and decay, and as a result there could be some topological suppression of the near-Skyrmion-to-vacuum transition. The chiral bag model is exactly a Skyrmion with a hole, so it is an interesting laboratory to study proton decay.  Does the intuitive picture or proton decay in the chiral bag model hold up? How large of a suppression is there? How sensitive is it to the bag radius? These are the sort of questions we aim to address here.
 
The setup of this paper is the following: we introduce the basic facts of the chiral langrangian and the Skyrme solution in Sec.~\ref{sec:skyrme}, followed by an introduction to quark bag models (Sec.~\ref{sec:hybridbag}). Section~\ref{sec:instanton1} contains a simplified calculation of Skyrmion unwinding where we neglect the interior (bag) dynamics. This calculation is subsequently improved in Sec.~\ref{sec:inter1} with a detailed calculation of the bag energy in the presence of time-dependent boundary conditions. Numerical results are presented in Sec.~\ref{sec:instanton1} and Sec.~\ref{sec:inter1}. We end with a discussion of our results and some directions for future work.

\section{Chiral Lagrangian and Skyrme Solution}
\label{sec:skyrme}

The low-energy effective theory for QCD is given by a non-linear sigma model, describing pions as the Goldstone bosons from the spontaneous symmetry breaking of chiral symmetry. In our setup, we work with two massless quark flavors, giving rise to an exact chiral $SU(2)$ symmetry. The pion Lagrangian is
\begin{equation}
\Lag_\pi=\frac{f_\pi^2}{4} \tr\left[ \partial_\mu U^\dagger \partial^\mu U\right],
\end{equation}
where $U$ is the $SU(2)$ valued pion field. This is a four-dimensional non-linear sigma model, where the ground state is given by constant field that we are free to choose to be $U=1$.

 It is convenient to introduce $X_\mu=U \partial_\mu U^\dagger=-X_\mu^\dagger$ which is an element of the $\mathfrak{su}(2)$ Lie algebra. The total energy of a static field configuration is given by
\begin{equation}
E_0=\frac{f_\pi^2}{4} \int d^3x\,\tr\left[ X_i X_i^\dagger\right].
\end{equation}
Finite energy solutions must have $U\rightarrow \mathbf 1$ at infinity, hence finite energy solution can be compactified to maps of $S^3\rightarrow SU(2)=S^3$. These maps are classified by $\pi_3(S^3)=\mathbb{Z}$, so there exist topological non-trivial field configurations. However, via simple scaling arguments, Derrick's theorem~\cite{Derrick} shows that these non-trivial field configurations are unstable against scaling. Skyrme~\cite{Skyrme:1961vq, Skyrme:1962vh} added an extra term to the Lagrangian
\begin{equation}
\mathcal L=\mathcal L_{\pi} + \frac{1}{32e^2}\tr \left[X_\mu,X_\nu\right]^2,
\end{equation}
which is the unique, lowest dimensional, higher-order term that satisfies all the symmetry constraints and is second order in time derivatives. This term stabilizes the field against scaling, thus allowing for stable non-trivial solitons. 

For our purposes, it is convenient to Wick rotate and go to dimensionless variables. Specifically, we set $x^{\mu} \rightarrow R\, y^{\mu}$ where we have introduced the characteristic length scale $R=\frac1{2e f_\pi}$, so from now on everything will be dimensionless. With this, the (Euclidean)  Skyrme action becomes
\begin{equation}
S_E = -\frac{1}{4e^2} \int d^4y\,\left(\frac14 \tr[X_a X_a]+\frac18 \tr[X_a, X_b]^2\right).
\label{eq:theL}
\end{equation}

The symmetry group of this model is $SU(2)_L\times SU(2)_R$ where $U\rightarrow LUR^\dagger$, in perfect correspondence with the QCD symmetry group. The axial $U(1)$ symmetry is broken because $U$ is restricted to $SU(2)$, which corresponds to the anomalous breaking of $U(1)_A$ in QCD. However, the baryonic $U(1)_V$ symmetry is not present, as $U\rightarrow e^{i\phi}Ue^{-i\phi}=U$, so there is no Noether current associated with the baryon current. However this theory has an extra conserved current,
\begin{equation}
B^\mu=\frac{\eps^{\mu\nu\alpha\beta}}{24\pi^2}\tr\left[X_\nu X_\alpha X_\beta\right].
\end{equation}
This current is conserved identically without invoking the equation of motion. This current is purely topological and its conserved charge
\begin{equation}
B=\int d^3x B^0(x)
\label{eq_winding_number}
\end{equation}
counts the topological winding of the $U$ field. Identifying $B$ with baryon number, the solitonic solutions of the theory are interpreted as baryons. With this identification, baryon number is a topologically conserved quantity, so protons cannot decay. This property, and the extra stabilizing feature which it requires, makes it reasonable that proton decay may be suppressed more than naively expected, even outside the strict confines of the Skyrme model.

\begin{figure}[h!]
\centering
\includegraphics[width=3.3in]{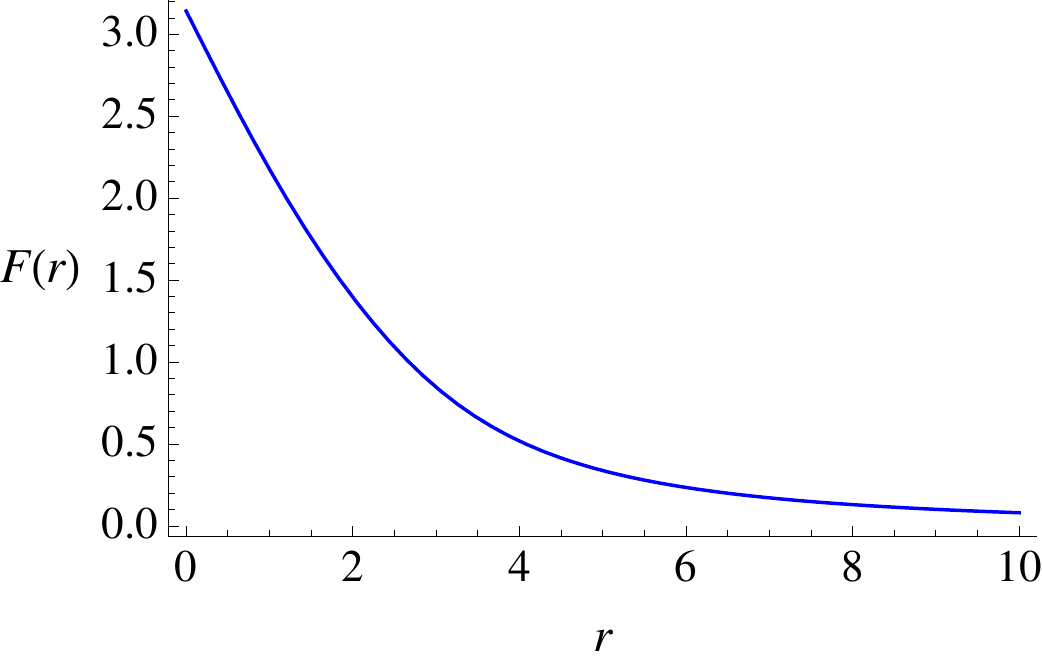}
\caption{Profile of the Skyrmion solution.}
\label{fig:skyrmeprof}
\end{figure}

The proton is the stable solitonic solution (Skyrmion) with winding number one. It can be found by making the {\em ansatz}
\begin{equation}
U=\exp\left[i\, F(r)\, \hat{x}\cdot\sigma\right]
\label{eq:ansatz}
\end{equation}
supplemented by the boundary conditions that $F(\infty)=0$ and $F(0)=\pi$. With these boundary conditions, $U$ wraps around the $SU(2)$ exactly once. For this ansatz, the energy is given by
\begin{align}
E[F] = & \frac{\pi}{2\,e^2}\int dr\, r^2\Big[\Big(F'^2+2\frac{\sin^2(F)}{r^2}\Big)+ \nonumber \\ 
 & \quad\quad \frac{4\,\sin^2(F)}{r^2}\Big(\frac{\sin^2(F)}{r^2}+2F'^2\Big)\Big].
 \label{eq:skyrmeq}
\end{align}
Minimizing $E[F]$ with respect to the function $F$ determines the profile of the Skyrmion, which is shown in Fig.~\ref{fig:skyrmeprof}.

The boundary condition at $r \rightarrow 0$ is set by the requirement that the Skyrmion has baryon number equal to one. Integrating the topological charge for the Skyrme solution from $r_0$ to infinity, we get
\begin{equation}
B=\frac1\pi(F(r_0)-\frac12 \sin 2 F(r_0)).
\label{eq:barysky}
\end{equation}
For $F(0) = \pi$, the baryon number $B = B(0) = 1$ .

Beautiful as this model is, experiments have clearly shown that the quark picture is the correct picture for small distance scales. The chiral bag model incorporates this by punching out a hole in the Skyrmion and replacing physics in that hole with a bag of free quarks; in essence, the hybrid bag models factorize QCD in short- and long-distance regimes. We will return later to how the hole effects Skyrmion properties. However, first we will review some properties of quark bag models.

\section{Chiral bag}
\label{sec:hybridbag}
As a first approximation, the inside of the bag simply contains free quarks. As such, it is described by the Minkowski path-integral
\begin{equation}
\int \D\bar\psi\D\psi \exp\Big\{i \int d^4x\, \bar\psi(i \slashed D)\psi \Big\},
\label{eq:pathint}
\end{equation}
where integration is restricted to a spherical region of radius $r_{bag}$. The boundary conditions, at $r_{bag}$, are chosen such that they respect the symmetries of QCD:
\begin{equation}
\begin{split}
i\,\slashed{n}\,\psi_L=U(n r_{bag})\psi_R.\\
i\,\slashed{n}\,\psi_R=U^\dagger(n r_{bag})\psi_L,\\
i\,\slashed{n}\,\psi=(U P_R+U^\dagger P_L) \psi=U_5 \psi
\end{split}
\label{eq:bagbc}
\end{equation}
Here $n_{\mu}$ is a radial unit vector pointing outwards, and the matrices $U$ are the non-linear pion field at the bag boundary. One can now easily see that this model has the right $SU(2)_L\times SU(2)_R$ symmetry structure, of $\psi_R\rightarrow R\psi_R,\,\psi_L\rightarrow L\psi_L,\, U\rightarrow LUR^\dagger$. For now we take
$U=\exp[i \theta n\cdot \sigma]$, $U_5=\exp[i \theta \gamma^5 n\cdot \sigma]$, for some value of $\theta$. However eventually we will make the identification $\theta = F(r_\text{bag})$, the solution to Eq.~(\ref{eq:skyrmeq}). This boundary condition will then link the properties inside the bag to the Skryme solution on the exterior. The conditions in Eq.~(\ref{eq:bagbc}) imply no vector currents flow through the boundary, as
\begin{equation}
-i\bar\psi\slashed{n}=i\psi^\dagger \slashed{n} \gamma^0=\psi^\dagger U_5^\dagger \gamma^0=\bar\psi U_5
\end{equation}
thus
\begin{equation}
i\bar\psi\slashed{n}\psi=\pm \bar\psi U_5\psi=0.
\end{equation}
Therefore, baryon and color currents through the bag boundary are zero\footnote{Because the boundary couples flavor states but leaves color untouched and we neglect full QCD inside, there is a $N_C$-fold degeneracy of all eigenstates. Therefore we focus on one color eigenstate and multiply by $N_C$ where appropriate in the rest of the paper.}. However, unlike the original (MIT) bag models, the axial $SU(2)$ current is continuous along the boundary of chiral bags once we identify $\theta = F(r)$, because of the full presence of the axial symmetry.

\begin{figure}[h!]
\centering
\includegraphics[width=2.75in, height=3.4in]{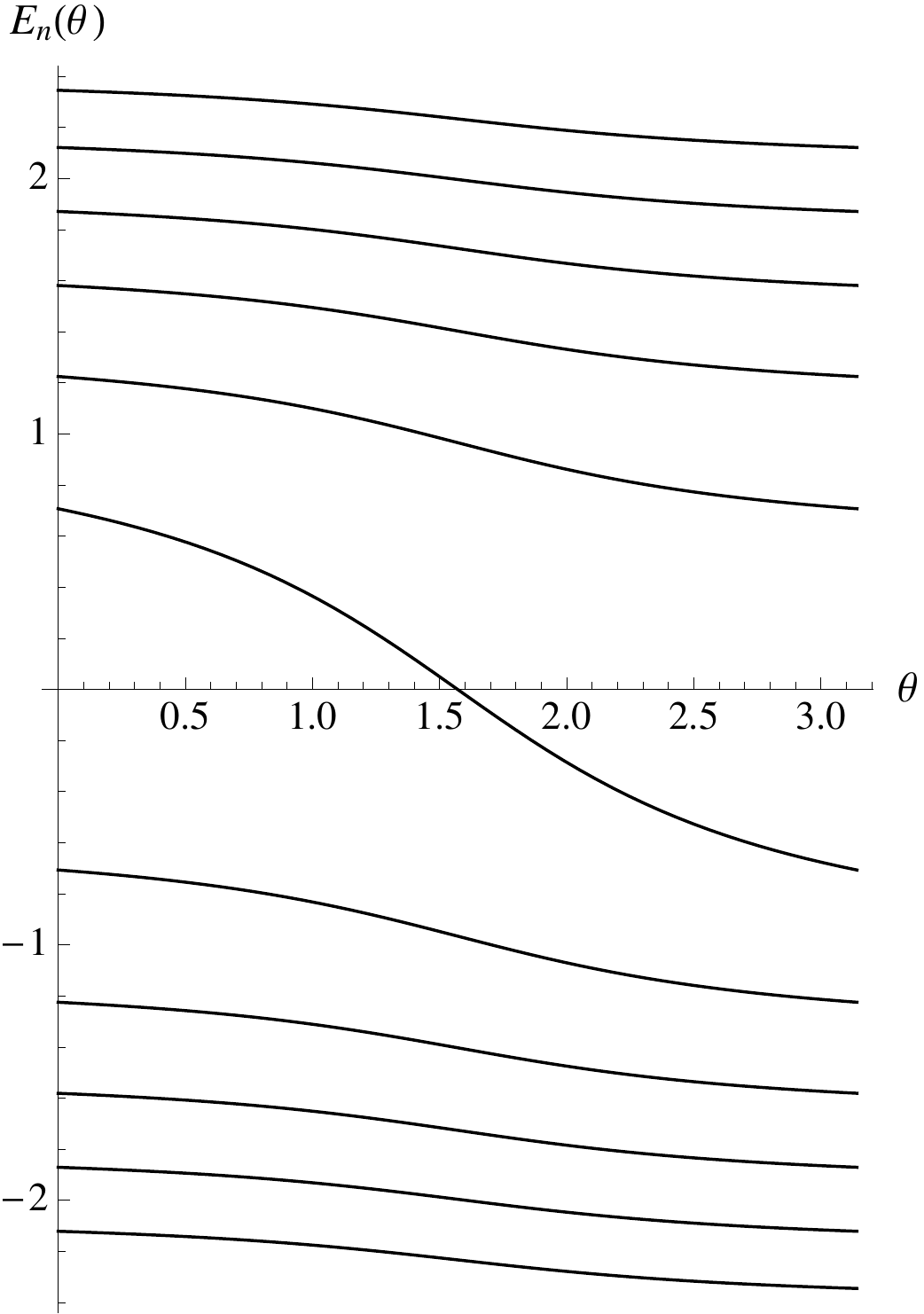}
\caption{Visualization of the eigenstates as function of $\theta$. The spectrum at $\theta=\pi$ is the same as $\theta=0$, but all levels go down by one step, $n \rightarrow n-1$. At $\theta=\pi/2$ there is a zero-mode.}
\label{fig:diracsea}
\end{figure}

We have shown that the baryon number is confined for fixed $\theta$. If $\theta$ varies, the baryon number of the Skyrmion changes, but so does the baryon number of the bag! The Dirac sea eigenstates energies are modified by and depend nontrivially on the boundary rotation $\theta$. Through the $n\cdot\sigma$ term on the boundary, spin and isospin are linked together, so eigenstates must be classified by the sum $\vec S + \vec I$, rather than spin or isospin individually. One can show using the map $\psi_E\rightarrow \gamma_5\psi_E$ that the spectrum is invariant under $E\rightarrow E$ and $\theta\rightarrow \theta+\pi$ hence periodic with period $\pi$. Using $\psi_E\rightarrow \gamma^0\psi_E$ once can show that the spectrum obeys the symmetry $E\rightarrow -E$ and $\theta\rightarrow \pi-\theta$. Therefore the spectrum has a zero-mode at $\theta=\pi/2$. The $\theta$-dependent eigenvalues have been worked out in Ref.~\cite{Mulders:1984df} and sketched in Fig.~\ref{fig:diracsea}. However, we are interested in properties of the bag as a whole, such as its total energy density or baryon number, rather than the individual eigenvalues.

The baryon number of the vacuum is defined with respect to a state in which every state is `half' filled. Every empty state therefore counts as baryon number $-\frac12$, and every filled state as $\frac12$. Summing over all modes, 
\begin{equation}
B_\text{sea}=-\frac12 \sum_n \text{sgn}(E_n),
\label{eq:barysea}
\end{equation}
which must be suitably regulated. In Ref.~\cite{Goldstone:1983tu}, the sum in Eq.~(\ref{eq:barysea}) was evaluated, and the baryon number for a chiral bag of radius $r$ was shown to be
\begin{equation}
B_\text{sea}=\frac1\pi\left(\left\{ \begin{array}{cc} -\theta & \theta < \pi/2 \\ \pi-\theta & \theta > \pi/2 \end{array} \right\}+\frac12\sin(2\theta(r))\right).
\label{eq:barybag}
\end{equation}
Identifying $\theta$ with the value of the Skyrme profile function at the boundary $F(r_{bag})$ and summing Eqs.~(\ref{eq:barybag}) and~(\ref{eq:barysky}), all dependence on $r_{bag}$ drops out and the net (Skyrmion + bag) baryon number remains constant at one.\footnote{Except when $\theta$ becomes less than $\pi/2$. We comment on this in section \ref{sec:inter1}.} This shows that, indeed, the topological current {\em is} the baryon current and baryon number is a globally conserved current. This remarkable property is referred to as the Cheshire Cat principle~\cite{Nadkarni:1984eg} in the literature and it gives confidence to the idea that (at least some properties of) baryons can be consistently factorized into short distance quark (and gluon) physics and long distance pion physics. In addition, the full $SU(2)_L \times SU(2)_R$ symmetry is maintained, making the chiral bag model a model for baryons.

Having seen that the quark bag and Skyrmion scenarios can be stitched together to form a consistent picture of the proton, we now want to study proton decay within this setup. We will proceed in steps. We first calculate the unwinding of a `punctured' Skyrmion neglecting changes to the bag interior. Then we add two crucial ingredients, the Casimir energy of the interior bag and source terms for baryon-number violation in the action. 

\section{Neglecting the interior}
\label{sec:instanton1}

As stated earlier, a punctured Skyrmion is not a topologically conserved state, so it can decay into the vacuum state.
Our approach to the classical solution describing the decay is the following: we take a Skyrmion with fixed puncture size $r_{bag}$, then consider dilations of that solution, $F(r) \rightarrow F(\lambda(t) r)$. The dilation parameter depends on time, with $\lambda(-\infty)=1$. As $\lambda$ grows, more of the Skyrmion is sucked into the hole, and as $\lambda \rightarrow \infty$ the Skyrmion disappears completely and we are left with the trivial ground state. Changing variables in Eq.~(\ref{eq:theL}) to $r\rightarrow \lambda(t) r$, and plugging in the Skyrme ansatz, we find
\begin{align}
&\mathcal S = \frac{\pi}{e^2} \int d\tau\int_{\lambda r_{bag}}^\infty dr\, \frac12\Big( \frac{r^4}{\lambda^5} + \frac{8\,r^2\,\sin^2(F)}{\lambda^3} \Big)\,F'^2\,\dot{\lambda}^2  \nonumber \\
&+\Big( \frac{2\,\sin^2(F) + r^2\,F'^2}{2\,\lambda} + \lambda\Big(\frac{4\,\sin^4(F)}{r^2} + 8\,\sin^2(F)\,F'^2\Big) \Big),
\label{eq:dilate}
\end{align}
where primes indicate derivatives with respect to $r$ and dots indicate a time derivative\footnote{As we have gone to dimensionless variables, time derivatives are taken with respect to $\tau = t/R$.}. Substituting the Skyrme solution $F(r, t)$ into the above and performing the spatial integration, we are left with a one-dimensional problem.
\begin{equation}
S[\lambda] = \int d\tau\, \dot{\lambda}^2\, K(\lambda) + V(\lambda).
\label{eq:lamlag}
\end{equation}
Because the integral for $K(\lambda)$ is convergent when one sets $r_{bag}=0$ in the lower limit, we have approximated $K(\lambda)$ by
\begin{equation}
K(\lambda)=\frac{A}{\lambda^3}+\frac{B}{\lambda^5}.
\label{eq:lambdadotmass}
\end{equation}
Plotting the potential $V(\lambda)$, there is a meta-stable minimum at $\lambda=1$, and an absolute minimum at $\lambda \rightarrow \infty$. Between the minima is the potential barrier associated with unwinding the Skyrmion. 
\begin{figure}[h!]
\centering
\includegraphics[width=3.0in]{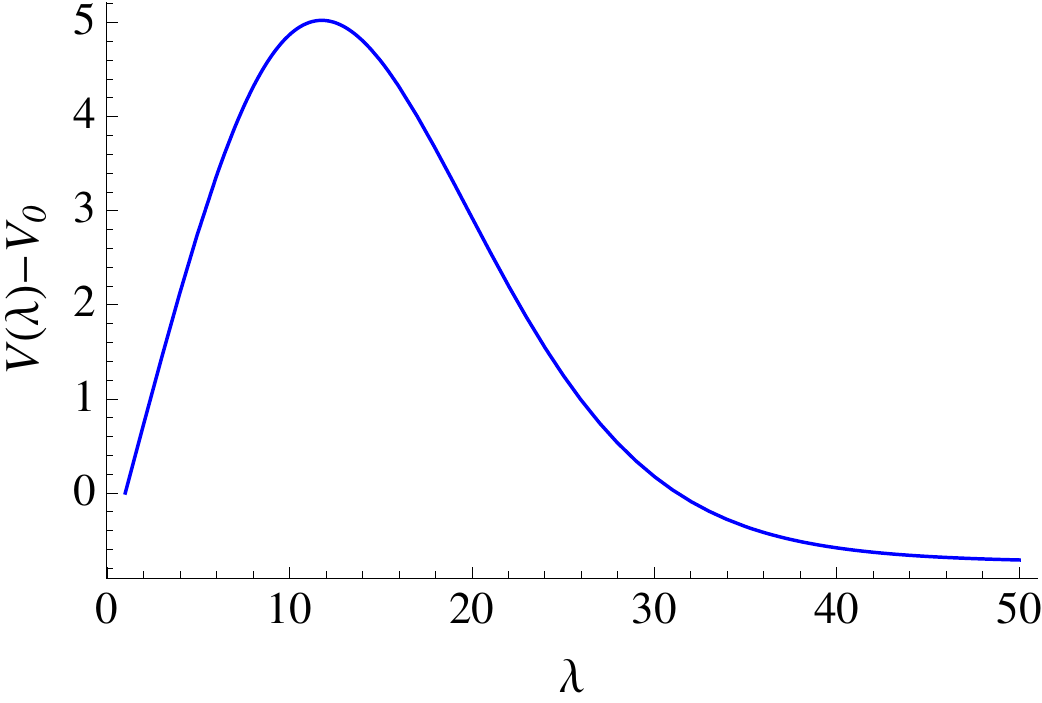}
\caption{The potential $V(\lambda)$ for a Skyrmion with bag radius $r_{bag} = 0.1\,R$, where we have subtracted off $V(1) \equiv V_0$. The tunneling probability is calculated by inverting the potential, $V \rightarrow - V$ and determining the bounce solution. For this particular example, the turn-around point is $\lambda^* \sim 28$.}
\label{fig:dilation}
\end{figure}
Transmission through the barrier can be calculated using instanton/WKB techniques, by which we construct a `bounce' solution that starts at $\lambda(-\infty)=1$, rolls down the inverted potential to the other side reverses, and ends up at $\lambda(+\infty)=1$  (in Eq.~(\ref{eq:dilate}), (\ref{eq:lamlag}) we have already been working with the Euclidean action). The
action of this `bounce' solution is\footnote{Note there has to be at local minimum at $\lambda=1$, unlike in Fig.\ref{fig:dilation}. However just imagine a small change to make it a local minimum. }
\begin{equation}
S_{tunnel} = 4\int_1^{\lambda^*}d\lambda\,\sqrt{K(\lambda)(V(\lambda) - V(1))}
\label{eq:tunnel}
\end{equation}
where $\lambda^*$ is the turnaround point for the bounce solution. The width of the decay of the punctured Skyrmion is proportional to $e^{-S_{tunnel}}$.  \\ 

The `bounce' solution can be thought of as the Skyrmion unwinding, then rewinding. To get an idea for the size of the suppression, we can plug in some numbers. For puncture sizes of $r_{bag}/R = 0.05, 0.1, 0.3$ we find exponential suppression of $5.7 \times 10^{-6},\,1.1 \times 10^{-4},\, 2.5\times 10^{-2}$ respectively.  From the height and width of the potential in Fig.~\ref{fig:dilation}, one may have expected a larger suppression. The reason the suppression is not larger is because of the $\lambda^{-3},\, \lambda^{-5}$ terms in $K(\lambda)$; these terms quickly shrink as $\lambda$ increases, resulting in a smaller than expected tunneling action.

 The calculation at this point clearly depends strongly on the size of the puncture, an issue we will return to soon. However, by neglecting the bag interior, this calculation is missing some physics. As the Skyrmion is pushed into the hole, the value $\theta=F(r_{bag}\,\lambda(\tau))$ of the Skyrmion at the bag boundary changes. The changing boundary condition shifts the energy levels of the bag fermions, an effect we need to incorporate. 

\section{Including the interior}
\label{sec:inter1}

\begin{figure}[h!]
\centering
\includegraphics[width=3.25in]{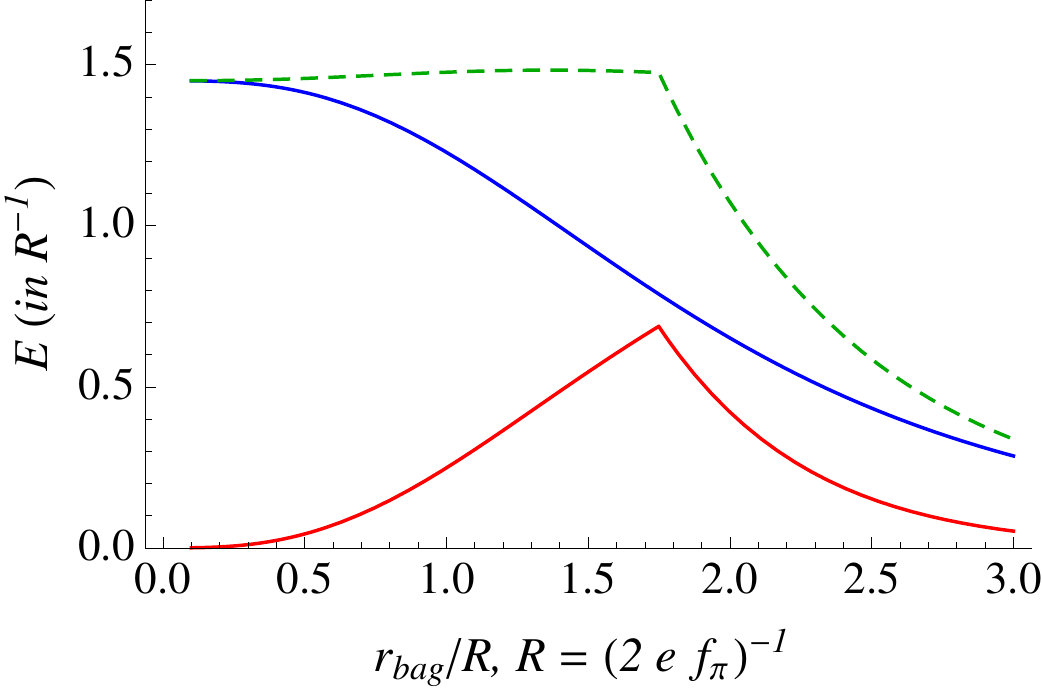}
\caption{The energy of the Skyrmion (dashed, in blue), the Casimir energy (solid, in red) and the total energy (dotted, in black) as function of the bag size.}
\label{fig:chiralbagenergy}
\end{figure}

To include the effects of the fermions in the bag we need to consider the effects of the fermionic path-integral. The resulting functional determinant, for static boundary conditions, was derived in Ref.~\cite{DeFrancia:1995xi} and is also presented in Appendix~\ref{sec:ECAS}. Schematically, the fermionic path integral results in a $\theta$-dependent functional determinant which, once regulated, can be shown to be $\exp[-T E_\text{cas}(\theta)]$, where $E_{cas}$ is the Casimir energy. Since it depends on $\theta$, the interior Casimir energy acts as an additional potential term in the action for $\theta$,
\begin{align}
&\int DU\D\psi^\dagger\D\psi \exp \Big\{-\int \mathcal\, d^4x \Big(L_E(\theta) + \psi^\dagger(\partial_\tau+H(\theta))\psi\Big) \Big\} \nonumber \\
& \rightarrow \int DU \exp \Big\{ -\int \mathcal\ d^4x\, L_E(\theta) + N_C\,\int d\tau\, E_{cas}(\theta) \Big\},
\label{eq:dpsi}
\end{align}
where the Hamiltonian is $H(\theta)=-i\gamma^0\slashed\nabla$ together with the boundary conditions (Eq.~\ref{eq:bagbc} with angle $\theta$). We have reverted back to using $\theta$ to describe the boundary angle in this section, however the reader should keep in mind that $\theta$ is set by the Skyrmion solution at the interface, $\theta = F(r_{bag})$. 
The Casimir energy of the spectrum of the Hamiltonian is defined as $E_{cas} = -\frac 1 2 \sum_n |E_n|$. We inserted a factor of $N_C$ to account for the color degeneracy. The total energy of a static solution is thus given by the sum of Skyrme energy (Eq.~(\ref{eq:skyrmeq})) and the Casimir energy (Eq.~(\ref{eq:thecasimir})). In Fig.~\ref{fig:chiralbagenergy}, the total energy is shown and one can see that it is remarkably flat as a function of the bag size, further evidence of Cheshire Cat principle.

The Casimir energy is~\cite{DeFrancia:1995xi}:
\begin{align}
& E_{cas}(\theta) = \frac{1}{r_{bag}}\,\Big(\,\frac{3}{4\pi}\,\Big(\left\{ \begin{array}{cc} \theta^2 & \theta < \pi/2 \\ (\pi-\theta)^2 & \theta > \pi/2 \end{array} \right\}  - \sin^2{\theta} \Big)  \nonumber \\
\label{eq:thecasimir} 
& ~~~~~~~~~~~~~~~~~~~~~~~ + C_2\, \sin^2(\theta) +  C_4\, \sin^4(\theta)  \\
& ~~~~~~~~~~~~~~~~~~~~~~~~~~~~~+ C_6\, \sin^6(\theta) + C_8\, \sin^8(\theta) \Big),\nonumber 
\end{align}
where the coefficients are:
\begin{align}
& C_2 = -0.13381,\quad C_4 = 0.05085 \\
\label{eq:thecasimir2} 
& C_6 = -0.01247,\quad C_8 = 0.01241.  \nonumber 
\end{align}

A few comments are in order regarding $E_{cas}(\theta)$. First, Eq.~(\ref{eq:thecasimir}) only contains the $\theta$ dependent pieces of the Casimir energy; $\theta$-independent terms, still proportional to $r_{bag}^{-1}$ do exist (see Ref.~\cite{DeFrancia:1995xi}) but do not effect our instanton calculation. Second, there is a sharp transition in $E_{cas}$ at $\theta = \pi/2$, which can be traced to the energy of the lowest lying eigenmode crossing zero. Varying $\theta$ from $< \pi/2$ to $> \pi/2$, a bag eigenmode is dragged out of the Dirac sea and becomes a valence mode. This state is filled but its energy (valence quarks) is not contained in the Casimir energy, so it must be added separately. Once included, the valence modes make the interior bag energy completely smooth in $\theta$. Likewise, the valence mode must be added separately to the baryon number of the Dirac sea (see Eq.~(\ref{eq:barybag})), keeping the net baryon number constant. 

Having reviewed the role of the bag action for static boundary conditions, we now need to see what happens when the boundary conditions change as a function of time. In our `bounce' solution describing Skyrmion unraveling, the boundary $F(r_{bag})$ changes as a function of time,  while, simultaneously, the bag dynamics will influence the optimal path of this `bounce' solution. As a first approximation, we assume the bag dynamics change adiabatically and simply change $E_{cas}(\theta) \rightarrow E_{cas}(F(r_{bag} \lambda(t) ) )$. In terms of the instanton calculation, this amount to changing the potential in Eq.~\ref{eq:lamlag}:
\begin{equation}
V(\lambda)\rightarrow V(\lambda)+E_{cas}(F(r_{bag}\,\lambda)).
 \label{eq:casplussky}
 \end{equation} 
 The new potential is shown below in Fig.~\ref{fig:chiralbagunstable} --the Casimir contribution is clearly important. We also point out that $E_{cas}$ depends on $r_{bag}$ in a relatively simple way, $E_{cas} \sim r^{-1}_{bag}$. 
\begin{figure}[h!]
\centering
\includegraphics[width=3.0in]{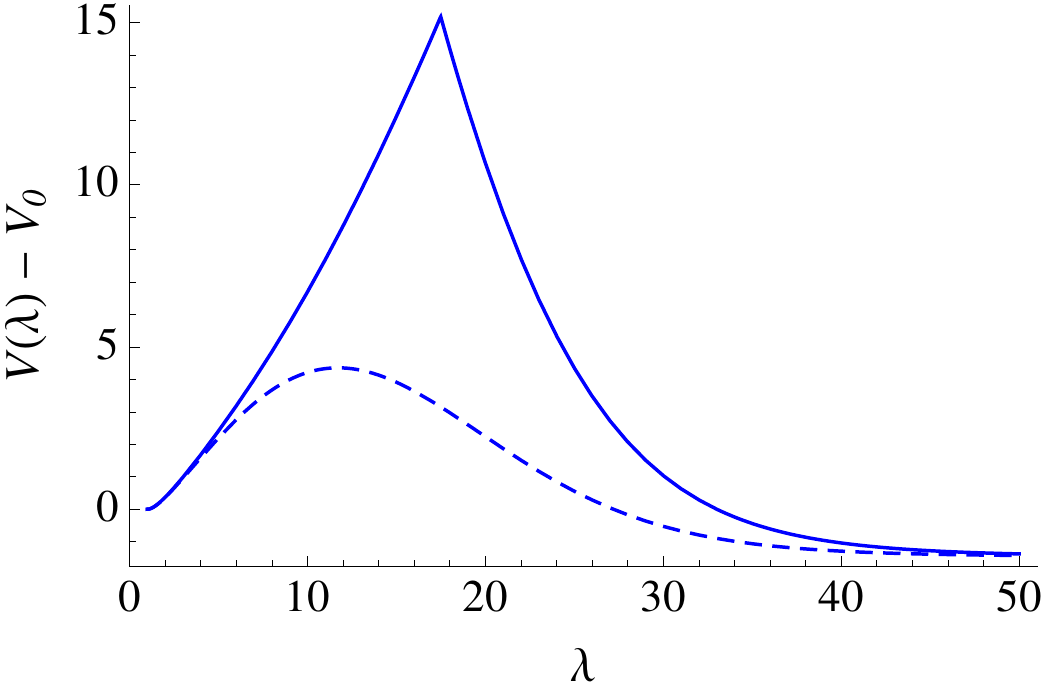}
\caption{The energy profile including the Casimir energy (solid). The Skyrmion contribution, as shown in Fig.~\ref{fig:dilation}, is indicated by the dashed line. }
\label{fig:chiralbagunstable}
\end{figure}
Using the Casimir-improved $V(\lambda)$ we calculate the tunneling exponential to be $6.7 \times 10^{-7}, 1.1\times 10^{-5}, 1.5\times 10^{-3}$ for $r_{bag}/R = 0.05, 0.1, 0.3$. Adding the Casimir has shrunk the tunneling by roughly an order of magnitude.

To see whether all consequences of time-dependent boundary conditions are captured by Eq ~(\ref{eq:casplussky}), a more thorough investigation is necessary. As the detailed calculation provided in the next section shows, Eq ~(\ref{eq:casplussky}) is not complete. An adiabatic calculation involves keeping the bag boundary static except for a short duration in which the bounce happens. Within the bounce, there are {\em two} special times which will prove to be important and whose effect is not included in Eq ~(\ref{eq:casplussky}):  at some time $t_1$ in the course of the unwinding, $F(r_{bag})$ will become less than $\pi/2$, signifying that a mode from the Dirac sea has been lifted out. Similarly, as the Skymion rewinds, this same mode will dive back into the sea at some later time $t_2$. This criss-crossing of $F(r_{bag}) = \pi/2$ indicates that  the system has a zero-mode, as shown in the cartoon of the Dirac sea in Fig.~\ref{fig:diracsea}. This zero-mode has important, subtle implications\footnote{By the Atiyah-Patodi-Singer theorem~\cite{Atiyah:1975jf, Atiyah:1976jg, Atiyah:1980jh}, if the system at any given instant has a zero mode solution, the full, time-dependent system will also exhibit a zero mode solution (see \cite{Nakahara:2003nw}).}. 

A second consequence of time-dependent boundary condition absent in Eq ~(\ref{eq:casplussky}) is that the Casimir energy depends on both $F(r_{bag})$ as well as on its time derivative, $\dot{F}(r_{bag})$. When manipulated into the bounce action, the $\dot F$ terms becomes $\dot{\lambda}$ terms, and the coefficient of the $\dot{\lambda}^2$ term will play the role of a `mass' for $\lambda$. As such, it  will affect the Skyrmion decay rate in a similar fashion to the $K(\lambda) $ term in Eq.~(\ref{eq:tunnel}).
 
\subsection{The non-static case}
\label{sec:thetadot}

\begin{figure}[h!]
\centering
\includegraphics[width=3.0in]{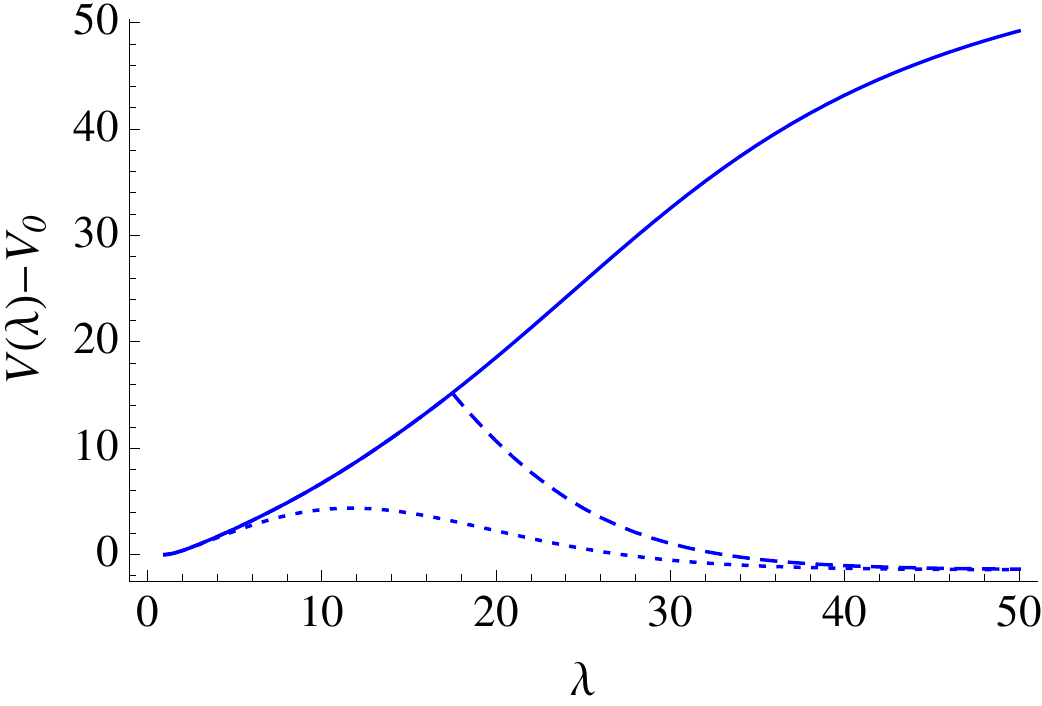}
\caption{The energy profile including the Casimir and the valence quarks. The Casimir (dashed) and Skyrmion (dotted) contributions are the same as in Fig.~\ref{fig:chiralbagunstable}.}
\label{fig:chiralbagstable}
\end{figure}

For time-dependent $\theta(\tau) = F(r_{bag}\lambda(\tau))$, the result of the fermionic path integral is $\det (\partial_\tau+H(\tau))$, where we have recast the time-dependent boundary conditions as a time-dependent Hamiltonian. In order to calculate the determinant we have to solve for the eigenvalues. Suppose
\begin{equation}
(\partial_\tau+H(\tau))|\psi(\tau)\rangle=\kappa\,|\psi(\tau)\rangle,
\end{equation}
and we define $\langle n(\tau)|$ to be the eigenstates of $H(\tau)$. We have
\begin{align}
& \langle n(\tau)|\frac{\partial}{\partial\tau}|\psi(\tau)\rangle+E_n(\tau)\langle n(\tau)|\psi(\tau)\rangle= \kappa\,\langle n(\tau)|\psi(\tau)\rangle \nonumber \\
& \frac{\partial}{\partial\tau}\langle n(\tau)|\psi(\tau)\rangle+E_n(\tau)\langle n(\tau)|\psi(\tau)\rangle- \\
& ~~~~~~~~~~~~~~~~~~~~~~~~~~~~~~~~~\langle \dot{n}(\tau)|\psi(\tau)\rangle =\kappa\langle n(\tau)|\psi(\tau)\rangle \nonumber
\end{align}
Defining $c_n(\tau)=\langle n(\tau)|\psi(\tau)\rangle$, we obtain
\begin{equation}
\dot{c}_n(\tau)+E_n(\tau) c_n(\tau)-\sum_m \langle\dot{n}(\tau)|m(\tau)\rangle c_m(\tau)=\kappa\,c_n(\tau).
\end{equation}
If the boundary conditions are changing slowly, the third term on the left-hand side is small and can be treated as a perturbation. We can rewrite the fermionic path integral as
\begin{multline}
\int \D c^\dagger\D c \exp\Big[-\int d\tau \big(c^\dagger_n(\tau) D_{nm}(\tau) c_m(\tau)-\\
c^\dagger_n(\tau) V_{nm}(\tau) c_m(\tau)\big)\Big],
\end{multline}
where $D_{nm}(\tau)=(\partial_\tau+E_n(\tau))\,\delta_{nm}$ and $V_{nm}(\tau)=\langle \dot{n}(\tau)|m(\tau)\rangle$.
Treating the first term in the exponent as the propagator and the second as a perturbation, the result is 
\begin{equation}
\det{\slashed D} = \det D_{nm} \exp\left[\sum\text{connected diagrams}\right]
\end{equation}
The determinant of $D_{nm}$ is easily evaluated because it is a disconnected set of one dimensional equations. The eigenfunctions $c_n(\tau)$ are
\begin{equation}
c_n(\tau)=\exp\left[\kappa \tau-\int^\tau_{-\frac{T}2} d\tau' E_n(\tau')\right].
\end{equation}
To determine $\kappa$, we impose anti-periodic temporal boundary conditions, $c_n(T/2) + c_n(-T/2) = 0$:
\begin{align}
 \kappa\, T\, - \int_{-\frac T 2}^\frac T 2\, d\tau'\, E_n(\tau') &=\, 2 \pi\, i\, \Big(m+\frac 1 2 \Big) \rightarrow \\
 \kappa = i \frac{2\pi(m+\frac 1 2)}{T} + \frac 1 T  \int_{-\frac T 2}^\frac T 2\ & d\tau'\, E_n(\tau') \equiv i\, \omega_m + \bar E_n, \nonumber \\
 \text{where} ~~ \bar E_n = & \frac 1 T \int_{-T/2}^{T/2} d\tau\, E_n(\tau)
 \label{eq:lambda}
\end{align}
In appendix \ref{sec:ECAS}  we provide an explicit calculation of $\det D_{nm}$, determined by the product over all $\kappa$. The result, for time-dependent boundary conditions, is
\begin{align}
\label{eq:tdepdet}
\det D_{nm} & = \exp[-T\,E_{cas} ] \\
T\, E_{cas} & =-\frac12 T\sum_n |\bar E_n|=-\frac12 \sum_n \left|\int d\tau E_n(\tau)\right|. \nonumber
\end{align}
Therefore, working to lowest order in an adiabatic approximation, we see that the functional determinant is the same as in the static case with the Casimir energy promoted to a function of time. Corrections to this result will be determined shortly, however, even at lowest order there is a subtlety incorporating Eq.~(\ref{eq:tdepdet}) into the action for $F(r_{bag})$. 
For all $n$ such that $E_n(\tau)$ doesn't change sign we have
\begin{equation}
\Big| \int\, d\tau E_n(\tau) \Big|= \int d\tau\, |E_n(\tau)|
\end{equation}
and there is no issue. However, as we noted regarding Eq.~(\ref{eq:thecasimir}), the sign of the lowest energy eigenvalue changes as we cross $F(r_{bag},\tau) = \theta = \pi/2$. In our case (the bounce), $E_0$ goes from negative to positive at $t_1$ and back at $t_2$, leading to
\begin{equation}
\Big| \int d\tau E_0(\tau) \Big|= \int d\tau |E_0(\tau)| - 2\int_{t_1}^{t_2}d\tau\, |E_0(\tau)|.
\end{equation}
Therefore the function determinant becomes
\begin{align}
\text{det} D_{nm} &= \exp{\Big\{-N_C\left(\int d\tau\, E_{cas}(\tau) + \int_{t_1}^{t_2}d\tau\, |E_0(\tau)|\right) \Big\} }, \nonumber \\
& ~~~~ E_{cas}(\tau)=-\frac12 \sum_n |E_n(\tau)|
\label{eq:withflip}
\end{align}
The extra piece in Eq.~(\ref{eq:withflip}) is exactly the valence term mentioned earlier. As we unwind the Skyrmion, we lift a negative state out of the sea and create the valence quarks. This means that we have yet another  extra potential term in our instanton barrier,
\begin{equation}
V(\theta)=E_{cas}(\theta)+\left\{ \begin{array}{cc} E_0(\theta) & \theta < \pi/2 \\ 0 & \theta > \pi/2 \end{array} \right\}.
\label{eq:casandvalpot}
\end{equation}
In Fig.~\ref{fig:chiralbagstable} the effect of this on the potential for $\kappa$ is shown. With the valence contribution, tunneling can no longer occur. The valence term makes the barrier in $\lambda$ impassable, and the chiral bag is a stable configuration. It appears Skyrmion unwinding is not possible, however the existence of the valence piece is intimately tied to the zero mode solution as we will show shortly \\.

\begin{figure}[!h]
\centering
\includegraphics[width=3.5in]{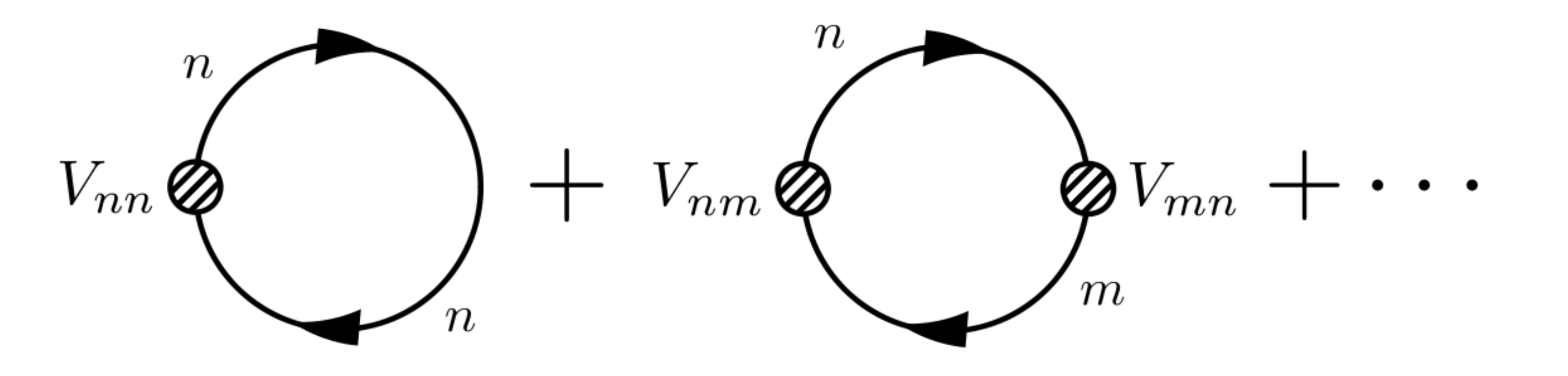}
\caption{The first few perturbations of the fermion functional determinant.}
\label{fig:diagrams}
\end{figure}

Before discussing the zero mode, we calculate the perturbative corrections to Eq.~(\ref{eq:withflip}). For this we need the propagator $D^{-1}_{nm}$. Focusing on one energy level, we define two functions
\begin{equation}
\begin{split}
\psi_{n,0}(\tau)=\exp\left[-\int_0^\tau d\tau' E_n(\tau')\right]\\
\tilde{\psi}_{n,0}(\tau)=\exp\left[\int_0^\tau d\tau' E_n(\tau')\right].\\
\end{split}
\end{equation}
These functions satisfy $D_{nm}\psi_{n,0}=\tilde{\psi}_{n,0} D_{nm}=0$. If $E_n(\pm\infty)<0$ then we can use $\psi_{n,0}, \bar{\psi}_{n,0}$ to construct the states
\begin{equation}
\begin{split}
\psi_{n,\tau'}(\tau)=\psi_{n,0}(\tau)\theta(\tau'-\tau)\\
\tilde{\psi}_{n,\tau'}(\tau)=\tilde{\psi}_{n,0}(\tau)\theta(\tau-\tau').\\
\end{split}
\end{equation}
These are normalizable functions for which it holds that
\begin{equation}
\begin{split}
D_{nm}\psi_{n,\tau'}(\tau)=-\psi_{n,0}(\tau')\delta(\tau-\tau')\\
\tilde{\psi}_{n,\tau'}(\tau)D_{nm}=-\tilde{\psi}_{n,0}(\tau')\delta(\tau-\tau')
\end{split}
\end{equation}
The inverse is solved through the following trick
\begin{equation}
\begin{split}
\tilde{\psi}_{m,\tau_1} D D^{-1} D\psi_{n,\tau_2}=\delta_{nm}\,\tilde{\psi}_0(\tau_1)\psi_0(\tau_2) \langle\tau_1|D^{-1}|\tau_2\rangle\\
\tilde{\psi}_{m,\tau_1} D \psi_{n,\tau_2}=-\delta_{nm}\,\tilde{\psi}_0(\tau_2)\psi_0(\tau_2)\theta(\tau_2-\tau_1)
\end{split}
\end{equation}
From which it follows
\begin{equation}
\langle\tau_1|D^{-1}_{nm}|\tau_2\rangle=-\delta_{nm}\,\exp\left[\int_{\tau_1}^{\tau_2} d\tau E_n(\tau)\right]\theta(\tau_2-\tau_1).
\label{eq:propagator1}
\end{equation}
Similarily for positive energy levels we have
\begin{equation}
\langle\tau_1|D^{-1}_{nm}|\tau_2\rangle=\delta_{nm}\,\exp\left[-\int_{\tau_2}^{\tau_1} d\tau E_n(\tau)\right]\theta(\tau_1-\tau_2).
\end{equation}
Now we can systematically include the corrections by calculating Feynman diagrams order by order in insertions of $V_{nm}$. Because the vertex is of degree 2, the
only connected Feynman diagrams are simple loops with $n$ vertices, as shown in Figure.~\ref{fig:diagrams}.
The loop with one vertex is zero, as we can always use rephasing freedom to set $\langle \dot{n}|n\rangle=0$. The second order correction is given by
\begin{equation}
-\frac12 \int d\tau d\tau' \sum_{n\neq m}V_{nm}(\tau) V_{mn}(\tau') D_{nm}(\tau', \tau) D_{mn}(\tau, \tau').\nonumber
\end{equation}
Due to the presence of $\theta$-functions in the propagators, the diagram is nonzero only if the energy levels $n$ and $m$ have opposite signs. Combining propagators (for $E_n > 0$), we get
\begin{equation}
D_{nm}(\tau', \tau) D_{mn}(\tau, \tau') \sim \exp\Big\{ -\int_{\tau}^{\tau'} d\xi\, (|E_n(\xi)| + |E_m(\xi)|) \Big\}.
\end{equation}
This propagator falls off exponentially in $|\tau-\tau'|$ because of the difference in energy between $n$ and $m$. Therefore, to a good approximation, we can just set $\tau' \rightarrow \tau$. Expanding $V_{mn}(\tau')$ about $\tau$ and keeping only the leading piece, the energy correction becomes
\begin{equation}
\frac12 \int d\tau \sum_{n,m}\frac{|V_{nm}(\tau)|^2}{|E_n(\tau)|+|E_m(\tau)|}.
\label{eq:adiabcorr}
\end{equation}
Using the chain rule, we have $\dot{n}=n'\dot{\theta}$. The energy correction above therefore adds to the coefficient of the $\dot{\theta}^2$ term which, in analogy with Eq.~(\ref{eq:lamlag}),  corresponds to the `mass' term of the instanton. 

Is this term ever important? Direct evaluation of  Eq.~(\ref{eq:adiabcorr}) is arduous and beyond the scope of this work, however we can get some intuition from dimensional analysis. From Eq.~(\ref{eq:adiabcorr}) we see that the non-adiabatic correction is proportional to $r_{bag}$, while $V(\lambda)$ is proportional to $r^{-1}_{bag}$. For bigger bag sizes, the potential barrier becomes smaller, but the smaller potential is compensated by a bigger mass term. The effects of the non-adiabatic piece are exacerbated because the Skyrme contribution to $K(\lambda)$ ($\sim \lambda^{-3})$ becomes vanishingly small at large $\lambda$ -- so any $\lambda$-independent (or mildly $\lambda$-dependent) piece of Eq.~(\ref{eq:adiabcorr}) will quickly become the dominant term. If $K(\lambda)$ goes to a constant, $V(\lambda)$ alone will determine the tunneling rate and the suppression will get bigger. However, without a full calculation of Eq.~(\ref{eq:adiabcorr}) we cannot be more quantitative. One option is to forget about the non-adiabatic corrections to $K(\lambda)$ and only use the Skyrmion contribution. This simplification limits us to small bag sizes, and really only gives us a lower limit on the size of the exponent.  As a  second option, we can parameterize the non-adiabatic contribution,
\begin{equation}
K(\lambda) \rightarrow  K(\lambda) + \text{const}\times  r_{bag}.
\end{equation}
To give a rough estimate, if the constant above is simply one, the exponential suppression becomes
\begin{equation}
e^{-S_{tunnel}} \sim \left\{ \begin{array}{cc} 10^{-35} & r_{bag}/R = 0.05 \\  2\times 10^{-16} & r_{bag}/R = 0.1 \\ 2\times10^{-5} & r_{bag}/R = 0.3 \end{array} \right.
\label{eq:nonadiabatapprox}
\end{equation}
However, before these numbers can be relevant, we need to somehow remove the stability exhibited in Fig.~\ref{fig:chiralbagstable}.

\subsection{The zero mode}
\label{sec:zeromode}

From Eq.~(\ref{eq:casandvalpot}) and Figure~\ref{fig:chiralbagstable} it is clear that the $E_0$ term from the valence quarks is what provides stability to the chiral bag system. The larger $\int_{t_1}^{t_2} d\tau |E_0(\tau)|$ is, the smaller the functional determinant and the larger the barrier inhibiting Skyrmion decay. How can Skyrmion decay occur in the face of this barrier? The answer lies in recalling that zero modes (or approximate zero modes) also lead to suppressed functional determinants.

 In a system with zero modes,  despite the fact that the functional determinant is vanishing, it is well known that correlations functions of operators can be nonzero. This fact is most easily proven by adding sources to the fermion Lagrangian, as functional derivatives with respect to the sources cancel the zeros in the determinant. To account for these non-vanishing correlation functions, a semi-local operator -- the 't Hooft vertex~\cite{ 'tHooft:1976up,'tHooft:1976fv} -- must be added to the action. 

Proceeding by analogy, when $\int_{t_1}^{t_2} d\tau |E_0(\tau)|$ is large there are approximate left and right zero modes of $D$. The existence of the zero modes makes the effective Skyrmion action zero unless sources are added. Including sources, the result of the fermionic path-integral $\mathcal Z[J, J^{\dag}]$ is
\begin{equation}
\exp\left[-N_C\,\left(\int d\tau E_\text{cas}(\tau) + \int_{t_1}^{t_2} d\tau |E_0(\tau)|\right)+J^\dagger D_0^{-1} J\right] 
\label{eq:zwithsource}
\end{equation}
From Eq.~(\ref{eq:propagator1}), we can see that functional derivatives with respect to the sources are accompanied by $D_0^{-1}$, which has exactly the right exponent to cancel the (otherwise large) $\int_{t_1}^{t_2} d\tau |E_0(\tau)|$ piece. Stated another way, in order for the Skyrmion to unwind, there {\em must} be some (microscopic) source for baryon-number violation present, such as the higher-dimensional operator
\begin{equation}
\frac{\epsilon^{ij...N_C}}{M^{3/2(N_C+1)-4}}\psi_i\psi_j ..\psi_{N_C} \chi \equiv \Lambda\, \epsilon^{ij...N_C} \psi_i\psi_j...\psi_{N_C}.
\label{eq:pdecay}
\end{equation}
Here $\chi$ represents some lepton and $M$ is the scale suppressing these operators; for convenience, we combine everything into a (fermionic) coefficient $\Lambda$. The need for these operators is no surprise -- we wouldn't be considering proton decay at all if they were forbidden!  To see how the operator in Eq.~(\ref{eq:pdecay}) generates the $D^{-1}_0$ needed to cancel the $E_0$ term, rewrite the interacting theory as
\begin{align}
 \exp\Big\{-\int \Big(\Lambda \Big(\frac{\delta}{\delta J}\Big)^{N_C} + \Lambda^{\dag}\Big( \frac{\delta}{\delta J^{\dag}} \Big)^{N_C}\Big) \Big\}\mathcal Z[J, J^{\dag}].
 \label{eq:lintbdecay}
\end{align}
Expanding out $\mathcal Z[J, J^{\dag}]$, the only nonzero term has $N_C$ of both $J$ and $J^{\dag}$, and therefore comes with $N_C$ copies of $D^{-1}_0$ -- just the right amount to cancel the $E_0$ term in the functional determinant. The relevant term being
\begin{equation}
\mathcal Z[J, J^{\dag}] \ni \exp\left[- N_C \int dt E_{cas}(t)\right] (J^\dagger(t_1) J(t_2))^{N_C}.
\end{equation}
The $J^\dagger$ destroys a particle at $t_1$ --  when its energy becomes positive -- and $J$ creates it when its energy becomes negative. The fermionic sources are cancelled when we apply the functional derivatives (in Eq.~(\ref{eq:lintbdecay})) corresponding the baryon number violating interaction. 

The resulting picture of proton decay becomes thus of a meta-stable chiral bag with only the negative states filled. There is an instanton bounce solution, where the Skyrmion unwinds pulling the valence quarks out of the vacuum at which point they are destroyed by the baryon violating operator. The bounce solution can be calculated by adding the Casimir energy to the potential, as postulated in the previous section. The bounce  comes with the prefactor $\Lambda^\dag \Lambda$. Therefore, the 
width of the proton is given by
\begin{equation}
\Gamma_p=\Lambda \Lambda^{\dag} e^{- S_{tunnel}}.
\end{equation}
The $\Lambda \Lambda^{\dag}$ gives the precise perturbative suppression of Eq.~\ref{eq:pertmatelem} as it should. We find, however, the extra {\em exponential} suppression of the hadronic matrix element coming from the unwinding. In the above, we have ruthlessly suppressed any other factor. The calculation as it stands will not give a precision result, however the presence of the exponential suppression in the hadronic matrix element is our main result, and it can be very relevant.

\section{Discussion}
\label{eq:disc}

In this work we have provided a rigorous instanton calculation for the decay of a ``baryon" in the context of the chiral bag model. The chiral bag model presents a model of the baryon where there are no valence quarks. It is a model where the valence quarks attract anti-quarks from the surrounding Dirac sea like a charge in a di-electric. In doing this it
generates a topological twist in the chiral phase of the condensate, described by the Skyrmion of the pion field. The binding of the valence quarks with
the anti-quarks lowers their energy, driving them into the Dirac sea of negative states. If this is an accurate description of the proton a non-perturbative
suppression is to be expected. Removal of the valence quarks leaves a state with higher energy and therefore it must proceed through a tunneling process.

The calculation we have done remains sensitive to the bag radius, and therefore does not obey the `Cheshire Cat principle' as nicely as one may have hoped. The chiral bag is, however,  just a leading order factorization model of hadronic QCD: the non-interacting free quark theory within the bag cannot possibly generate the spontaneous symmetry breaking necessary for the pion field, so the bag size cannot be arbitrarily large. Similarly, the pion field description also breaks down for small bag size, due to the running of the coupling. Given the limitations of the chiral bag, dependence on $r_{bag}$ it not surprising. Numbers aside, our main qualitative point is that this model gives
the right picture of what a proton is, and that therefore a tunneling suppression is to be expected. If a more sophisticated calculation including NLO effects results in a regime in which the calculation stabilizes, this would certainly strengthen our result.

Applying the suppression we find to simple GUT models has profound implications. An additional $\sim10^{-4}$ suppression would mean the unification scale could be lowered by 
an order of magnitude for dimension-six baryon-number violation, or two orders of magnitude for dimension-five baryon number violation. Furthermore, $10^{-4}$ should be viewed as an {\em upper} limit. Including a parametrization of the non-adiabatic terms (see Eq.~(\ref{eq:nonadiabatapprox})), we found $10^{-12}$ suppression was perfectly reasonable, implying a three (six) order of magnitude drop for dimension six (five) baryon number violation. A more exact value for the exponent requires a rigorous calculation of  Eq.~(\ref{eq:adiabcorr}). A change in the required GUT scale of this order would certainly resuscitate several scenarios~\cite{Ellis:1981tv, Nath:1985ub, Nath:1988tx, Hisano:1992jj, Murayama:2001ur, Dutta:2004zh, Fukuyama:2004xs}!  According to our calculation other tests of baryon number violation, such as neutron-antineutron oscillation~\cite{ Kuzmin:1970nx, Mohapatra:1980qe, Chang:1980ey}, should also be highly suppressed.

Throughout our calculations we have completely ignored any mass for the interior quarks. Adding in a mass, the energy levels shift, and if the shift is large enough the mass can effectively prevent valence quarks from diving in the vacuum. If the valence quarks retain positive energy, they can decay immediately and do not need to be lifted by unwinding the pion field. The baryon is then an ordinary bound state. The role quark masses play in this calculation may also explain the apparent disagreement between our result and estimations of proton decay matrix elements based on lattice QCD~\cite{Tsutsui:2004qc, Aoki:2008ku, Cooney}. The suppression in the chiral bag model comes primarily from the chiral symmetry and its twisting. Exact chiral symmetry is a difficult regime to probe on the lattice due to difficulties with chiral symmetry of fermions and with fitting the Compton wavelength of low mass states into the finite lattice volume. Lattice results therefore depend on extrapolation of quark mass and system volume into the physical regime. If the dependence of proton decay matrix elements on lattice artifacts is different than for the more conventional observables (meson masses, etc.), extrapolations, motivated by conventional observables, would be inappropriate and may explain the apparent differences between our result and the lattice. The mass-dependence in our result is a subtle issue and deserving of additional study.

\appendix
\section{Deriving the Casimir energy}
\label{sec:ECAS}

The eigenvalue spectrum is given by
\begin{equation}
\kappa=i\left(\frac{2\pi\left(m+\frac12\right)}{T}\right)+\bar E_n
\end{equation}
This spectrum is symmetrical around the real axis, so we can pair every conjugate pair and we have
\begin{equation}
\det D=\prod_n \prod_m (\omega_m^2+\bar E_n^2)^{\frac12}
\end{equation}
Using zeta-function regularization we have
\begin{equation}
\det D=\exp\left[-\frac12\frac{d}{ds} \sum_n \sum_m (\omega_m^2+\bar E_n^2)^{-s}\right]
\end{equation}
In the limit $T\rightarrow\infty$ we can replace $\sum_m\rightarrow \frac{T}{2\pi} \int d\omega$,
we obtain
\begin{equation}
\int d\omega (\omega^2+\bar E_n^2)^{-s}=|\bar E_n|^{-2s+1} \sqrt{\pi} \Gamma(s-\frac12)/\Gamma(s)
\end{equation}
Because of the $\Gamma[s]=\Gamma[s+1]/s$ we see that the above expression is 0 at $s=0$. Therefor
the only way to get a non-zero values is if the differential operator is applied to kill $s$. We are thus left
with
\begin{equation}
\det D=\exp[-\frac12 \sum_n |\bar E_n|^{-2s+1} \frac{T}{2\pi} \sqrt{\pi} \Gamma(s-\frac12)/\Gamma(s+1)]|_{s=0}.
\end{equation}
This can be written as
\begin{equation}
\det D=\exp[- E_{cas} T ],
\end{equation}
where the Casimir energy $E_\text{cas}$ is
\begin{equation}
E_{cas}=-\frac12 \Big(\sum |\bar E_n| (\bar E_n^2)^{-s}\Big) \Big|_{s=0},
\end{equation}
--  exactly the zeta-function regularization of the vacuum energy.
                         
\section*{Acknowledgments}

We thank Paddy Fox, Roni Harnik, Chris Hill, Andreas Kronfeld and Ethan Neil for many valuable conversations and acknowledge Joachim Kopp for collaboration during the early stages of this work. GS acknowledge Piet Mulders for a useful discussion. AM and GS are supported by Fermilab operated by Fermi Research Alliance, 
LLC under contract number DE-AC02-07CH11359 with the 
US Department of Energy.


\begin{thebibliography}{99}

\bibitem{Georgi:1974sy}
  H.~Georgi, S.~L.~Glashow,
  Phys.\ Rev.\ Lett.\  {\bf 32}, 438-441 (1974).
  

\bibitem{Tsutsui:2004qc}
  N.~Tsutsui {\it et al.}  [CP-PACS Collaboration and JLQCD Collaborations],
  Phys.\ Rev.\  D {\bf 70}, 111501 (2004)
  [arXiv:hep-lat/0402026].
  
\bibitem{Aoki:2008ku}
  Y.~Aoki {\it et al.}  [RBC-UKQCD Collaboration],
  Phys.\ Rev.\  D {\bf 78}, 054505 (2008)
  [arXiv:0806.1031 [hep-lat]].

\bibitem{Cooney}
P.~Cooney, 
http://www.era.lib.ed.ac.uk/handle/1842/4042

\bibitem{Claudson:1981gh}
  M.~Claudson, M.~B.~Wise, L.~J.~Hall,
  Nucl.\ Phys.\  {\bf B195}, 297 (1982).
  

\bibitem{Jackiw:1975fn}
  R.~Jackiw, C.~Rebbi,
  Phys.\ Rev.\  {\bf D13}, 3398-3409 (1976).

\bibitem{Vento:1980mu}
  V.~Vento, M.~Rho, E.~M.~Nyman, J.~H.~Jun, G.~E.~Brown,
  Nucl.\ Phys.\  {\bf A345}, 413 (1980).
  
\bibitem{Goldstone:1981kk}
  J.~Goldstone, F.~Wilczek,
  Phys.\ Rev.\ Lett.\  {\bf 47}, 986-989 (1981).
  
\bibitem{Goldstone:1983tu}
  J.~Goldstone, R.~L.~Jaffe,
  Phys.\ Rev.\ Lett.\  {\bf 51}, 1518 (1983).

\bibitem{Brown:1984sx}
  G.~E.~Brown, A.~D.~Jackson, M.~Rho, V.~Vento,
  Phys.\ Lett.\  {\bf B140}, 285-289 (1984).

\bibitem{Mulders:1984df}
  P.~J.~Mulders,
  Phys.\ Rev.\  {\bf D30}, 1073 (1984).
  
\bibitem{Vepstas:1984sw}
  L.~Vepstas, A.~D.~Jackson, A.~S.~Goldhaber,
  Phys.\ Lett.\  {\bf B140}, 280-284 (1984).
  
  
\bibitem{Vepstas:1990tw}
  L.~Vepstas, A.~D.~Jackson,
  Phys.\ Rept.\  {\bf 187}, 109-143 (1990).
  

\bibitem{Skyrme:1961vq}
  T.~H.~R.~Skyrme,
  Proc.\ Roy.\ Soc.\ Lond.\  {\bf A260}, 127-138 (1961).
  
\bibitem{Skyrme:1962vh}
  T.~H.~R.~Skyrme,
  Nucl.\ Phys.\  {\bf 31}, 556-569 (1962).   
    
\bibitem{Chodos:1974je}
  A.~Chodos, R.~L.~Jaffe, K.~Johnson, C.~B.~Thorn, V.~F.~Weisskopf,
  Phys.\ Rev.\  {\bf D9}, 3471-3495 (1974).
  
\bibitem{DeGrand:1975cf}
  T.~A.~DeGrand, R.~L.~Jaffe, K.~Johnson, J.~E.~Kiskis,
  Phys.\ Rev.\  {\bf D12}, 2060 (1975).

\bibitem{Inoue:1975bk}
  T.~Inoue, T.~Maskawa,
  Prog.\ Theor.\ Phys.\  {\bf 54}, 1833 (1975).
        
\bibitem{Callan:1978bm}
  C.~G.~Callan, Jr., R.~F.~Dashen, D.~J.~Gross,
  Phys.\ Rev.\  {\bf D19}, 1826 (1979).
  
\bibitem{Brown:1979ui}
  G.~E.~Brown, M.~Rho,
  Phys.\ Lett.\  {\bf B82}, 177-180 (1979).

\bibitem{Brown:1979ij}
  G.~E.~Brown, M.~Rho, V.~Vento,
  Phys.\ Lett.\  {\bf B84}, 383 (1979).
  
\bibitem{Nadkarni:1984eg}
  S.~Nadkarni, H.~B.~Nielsen, I.~Zahed,
  Nucl.\ Phys.\  {\bf B253}, 308 (1985).
  
\bibitem{Rho:1994ni}
  M.~Rho,
  Phys.\ Rept.\  {\bf 240}, 1 (1994).
  [hep-ph/0206003].
  
\bibitem{DeFrancia:1995xi}
  M.~De Francia, H.~Falomir, E.~M.~Santangelo,
  Phys.\ Lett.\  {\bf B371}, 285-292 (1996).
  [hep-ph/9507347].
    
\bibitem{Witten:1983tx}
  E.~Witten,
  Nucl.\ Phys.\  {\bf B223}, 433-444 (1983).
  
\bibitem{Witten:1983tw}
  E.~Witten,
  Nucl.\ Phys.\  {\bf B223}, 422-432 (1983).

\bibitem{Derrick} 
  G.H. Derrick,
  J. Mathematical Phys. 5 (1964), pp. 1252Ð1254
  
\bibitem{'tHooft:1976fv}
  G.~'t Hooft,
  Phys.\ Rev.\  {\bf D14}, 3432-3450 (1976).
  
\bibitem{'tHooft:1976up}
  G.~'t Hooft,
  Phys.\ Rev.\ Lett.\  {\bf 37}, 8-11 (1976).
  
\bibitem{Atiyah:1975jf}
  M.~F.~Atiyah, V.~K.~Patodi, I.~M.~Singer,
  Math.\ Proc.\ Cambridge Phil.\ Soc.\  {\bf 77}, 43 (1975).
  
\bibitem{Atiyah:1976jg}
  M.~F.~Atiyah, V.~K.~Patodi, I.~M.~Singer,
  Math.\ Proc.\ Cambridge Phil.\ Soc.\  {\bf 78}, 405 (1976).

\bibitem{Atiyah:1980jh}
  M.~F.~Atiyah, V.~K.~Patodi, I.~M.~Singer,
  Math.\ Proc.\ Cambridge Phil.\ Soc.\  {\bf 79}, 71 (1980).   

\bibitem{Nakahara:2003nw}
 M.~Nakahara, 
 Geometry, topology and physics (2003)
  
\bibitem{Ellis:1981tv}
  J.~R.~Ellis, D.~V.~Nanopoulos, S.~Rudaz,
  Nucl.\ Phys.\  {\bf B202}, 43 (1982).
 
\bibitem{Nath:1985ub}
  P.~Nath, A.~H.~Chamseddine, R.~L.~Arnowitt,
  Phys.\ Rev.\  {\bf D32}, 2348-2358 (1985).
 
\bibitem{Nath:1988tx}
  P.~Nath, R.~L.~Arnowitt,
  Phys.\ Rev.\  {\bf D38}, 1479 (1988).
 
\bibitem{Hisano:1992jj}
  J.~Hisano, H.~Murayama, T.~Yanagida,
  Nucl.\ Phys.\  {\bf B402}, 46-84 (1993).
  [hep-ph/9207279].
    
\bibitem{Murayama:2001ur}
  H.~Murayama, A.~Pierce,
  Phys.\ Rev.\  {\bf D65}, 055009 (2002).
  [hep-ph/0108104].
  
\bibitem{Fukuyama:2004xs}
  T.~Fukuyama, A.~Ilakovac, T.~Kikuchi, S.~Meljanac, N.~Okada,
  Eur.\ Phys.\ J.\  {\bf C42}, 191-203 (2005).
  [hep-ph/0401213].
  
\bibitem{Dutta:2004zh}
  B.~Dutta, Y.~Mimura, R.~N.~Mohapatra,
  Phys.\ Rev.\ Lett.\  {\bf 94}, 091804 (2005).
  [hep-ph/0412105].
  
\bibitem{Kuzmin:1970nx}
  V.~A.~Kuzmin,
  Pisma Zh.\ Eksp.\ Teor.\ Fiz.\  {\bf 12}, 335-337 (1970).
  
\bibitem{Mohapatra:1980qe}
  R.~N.~Mohapatra, R.~E.~Marshak,
  Phys.\ Rev.\ Lett.\  {\bf 44}, 1316-1319 (1980).
  
\bibitem{Chang:1980ey}
  L.~N.~Chang, N.~P.~Chang,
  Phys.\ Lett.\  {\bf B92}, 103-106 (1980).
  

\end{thebibliography}
\end{document}